\documentclass[12pt]{iopart}

\def\mn{{\mu\nu}}
\def\mnab{{\mu\nu\al\be}}

\def\ep{\epsilon}
\def\et{\eta}
\def\al{\alpha}
\def\be{\beta}
\def\ga{\gamma}

\def\ka{\kappa}
\def\la{\lambda}

\def\si{\sigma}
\def\ta{\tau}
\def\ve{\varepsilon}
\def\ps{\psi}
\def\ch{\chi}

\def\bt{{\tilde b}}
\def\ct{{\tilde c}}
\def\dt{{\tilde d}}
\def\gt{{\tilde g}}
\def\Ht{{\tilde H}}

\def\cl{{\cal L}}
\def\ol#1{\overline{#1}}
\def\Ga{\Gamma}
\def\lrvec#1{ \stackrel{\leftrightarrow}{#1} }
\def\prt{\partial}

\def\fr#1#2{{{#1} \over {#2}}}
\def\half{{\textstyle{1\over 2}}}
\def\frac#1#2{{\textstyle{{#1}\over {#2}}}}
\def\abs#1{\left|{#1}\right|}

\def\hacek{h\'{a}\v{c}ek}

\newcommand{\rf}[1]{(\ref{#1})}

\def\gev{\mbox{ GeV}}

\def\hlineone{\br}
\def\hlinetwo{\mr}
\def\hlinethree{\br}

\newcommand{\beq}{\begin{equation}}
\newcommand{\eeq}{\end{equation}}
\newcommand{\bea}{\begin{eqnarray}}
\newcommand{\eea}{\end{eqnarray}}

\def\gm#1#2#3{g^{(M)}_{#1#2#3}}
\def\gA#1{g^{(A){#1}}}
\def\glA#1{g^{(A)}_{#1}}
\def\ggA#1#2#3{g^{(A)}_{#1#2#3}}
\def\gT#1{g^{(T)}_{#1}}
\def\guT#1{g^{(T)#1}}
\def\ggT#1#2#3{g^{(T)}_{#1#2#3}}
\def\gg#1#2#3{g_{#1#2#3}}

\def\etal{{\it et al.}}

\begin{document}

\title{Fermion observables for Lorentz violation}
\author{A.\ Fittante and N.\ Russell}
\address{Physics Department,
Northern Michigan University,
Marquette, MI 49855, U.S.A.}

\begin{abstract}
The relationship between experimental observables for Lorentz violation
in the fermion sector
and the coefficients for Lorentz violation appearing in the
lagrangian density is investigated
in the minimal Standard-Model Extension.
The definitions of the $44$ fermion-sector observables,
called the tilde coefficients,
are shown to have a block structure.
The c coefficients decouple from all the others,
have six subspaces of dimension $1$,
and one of dimension $3$.
The remaining tilde coefficients form eight blocks,
one of dimension $6$,
one of dimension $2$, three of dimension $5$,
and three of dimension $4$.
By inverting these definitions,
thirteen limits on the electron-sector tilde coefficients
are deduced.
\end{abstract}
\maketitle

\section{Introduction}
The possibility of Lorentz violation in nature
has received much attention in recent years,
triggered by the pioneering work
of Kosteleck\'y, Samuel, and Potting
\cite{ksp}.
Numerous investigations of Lorentz symmetry have been carried
out in systems
involving ordinary fermionic matter:
electrons, protons, and neutrons.
In the case of electrons,
these have included studies with
spin-polarized torsion pendula
\cite{08HeckelPRD,TorsionPend},
atomic transitions \cite{2009Altschul2S1S},
Penning traps \cite{99Dehmelt,1999Mittleman},
colliders \cite{2010Altschullabe},
optical and microwave resonators
\cite{resonators},
and
astrophysical results \cite{astrophysics,07AltschulAstrophys}.
In fact,
Lorentz-symmetry investigations of a theoretical and experimental nature
cover all subfields of physics
\cite{cptprocs}.
Proton-based investigations have been performed
with comagnetometers
\cite{Romalis,1995berglund,99akcl},
the hydrogen maser
\cite{Hmaser},
Cesium fountain clocks
\cite{2006Wolf},
Doppler-shifted systems
\cite{doppler},
and the Penning trap
\cite{1999Gabrielse}.
Among the investigations done with neutrons
are ones with
clock-comparisons \cite{clockcomparisons},
magnetometers \cite{magnetom,09AltschulClocks,Romalis},
ultra-cold neutrons \cite{2009Altarev},
masers \cite{maser_n,04Cane}
and
astrophysical data \cite{Astrophs_n}.

In the general framework for Lorentz violation
known as the Standard-Model Extension, or SME,
the minimal theory has 44 independent experimental observables for each
of the three fermions making up ordinary matter.
As of January 2012,
experimental sensitivities exist
for about 58\% of them \cite{12Tables}.
In this work,
our goal is to study the structure of these observables,
known as the `tilde' coefficients,
and to seek relationships between them that may be
used to deduce limits from theoretical considerations.

The SME is a general realistic effective field theory for Lorentz violation
\cite{97SME, 98SME, akgrav},
providing for minuscule violations of CPT \cite{02greenberg}
and Lorentz symmetry
in the Standard Model
and General Relativity.
The framework is set up using a Lagrange density
containing conventional terms
supplemented with unconventional ones,
each a coordinate independent product
of a coefficient for Lorentz violation
and a Lorentz-breaking operator.
The operators
can be classified according to their mass dimension,
and the minimal SME
involves mass dimensions 3 and 4 only.
Apart from investigations in the
ordinary-matter fermionic sectors,
dozens of experiments have been conducted
to investigate whether any of the coefficients
of the Lorentz-violating operators are nonzero
\cite{12Tables}.

Studies of the minimal SME have found
the dispersion relation for a free fermion
in a constant background \cite{2001Lehnert},
ways to factorize it \cite{10quaternions},
and methods for deducing
the associated classical Lagrange function
\cite{classical}.
Other investigations have explored
the geometries relevant in curved spacetimes \cite{2011finsler},
and have looked at the nonminimal operators
in the photon sector,
where a classification exists for all mass
dimensions \cite{09Nonmin}.
Recent work has studied the nonminimal neutrino sector
\cite{12nu}.
Here,
we primarily consider the observables for
experiments with electrons, protons, and neutrons
\cite{99akcl,99akclNonRel,SpaceTests}
in which boost effects can be neglected.

\section{Minimal fermion sector in flat spacetime}
In the minimal SME,
the flat-space lagrangian density $\cl$ describing a spin-$\half$
fermion $\ps$ of mass $m$ is
\cite{97SME,98SME}
\beq
\cl = \frac{1}{2}i \ol{\ps} \Ga_\nu \lrvec{\prt^\nu} \ps
   - \ol{\ps} M \ps
\quad ,
\label{lagr}
\eeq
where
\beq
M := m + a_\mu \ga^\mu + b_\mu \ga_5 \ga^\mu
   + \half H_\mn \si^\mn
\quad
\label{M}
\eeq
and
\bea
\Ga_\nu &:=& \ga_\nu + c_\mn \ga^\mu + d_\mn \ga_5 \ga^\mu
+e_\nu + i f_\nu \ga_5 + \half g_{\la \mu \nu} \si^{\la \mu}
\quad .
\label{Gam}
\eea
The coefficients
$a_\mu$,
$b_\mu$,
and $H_{\mu\nu}$
appearing in \rf{M} have dimensions of mass
and control operators of mass dimension 3,
while the coefficients
$c_{\mu\nu}$,
$d_{\mu\nu}$,
$e_\mu$,
$f_\mu$,
and
$g_{\mu\nu\la}$
appearing in \rf{Gam}
are dimensionless
and control operators of mass dimension 4.
In the minimal SME,
there are three distinct fermionic sectors,
and the coefficients in each are denoted by superscripts:
$e$ for electrons and positrons,
$p$ for protons and antiprotons,
and
$n$ for neutrons and antineutrons.
When working in just one particle-antiparticle sector,
these superscripts are often suppressed.
Within one sector,
experiments with sensitivities to these coefficients
are likely to involve
comparisons of the particle with its antiparticle,
comparisons of differing spin states of one particle,
or comparisons of differing motional states of a particle.

By definition,
$c_{\mn}$ and $d_{\mn}$
are traceless,
$g_{\mu\nu\la}$ is antisymmetric in the first two indices,
and
$H_{\mu\nu}$ is antisymmetric.
This gives a total
of $4+4+15+15+4+4+24+6=76$
components of the coefficients
$a_\mu$ through $H_{\mu\nu}$,
respectively,
for a single fermion.

To allow comparison between results from different experiments,
measurements of the Lorentz-breaking background fields are
reported in the standard Sun-centered inertial reference frame,
with coordinates denoted by upper-case roman letters
$(T,X,Y,Z)$ \cite{SpaceTests}.
This has $Z$ axis parallel to the rotational axis of the Earth,
and $X$ axis pointing from the Sun towards the northern vernal equinox.

We assume a Minkowski spacetime throughout,
so that the metric is diagonal.
Where sign choices are necessary,
we choose spacetime metric $\et_{\mn}$ with $\et_{TT}=+1$,
and antisymmetric tensor $\ep^{\mnab}$ defined with $\ep^{TXYZ} = +1$.
In a few places, indices $J$, $K$, $L$,
can take on possible values
$X, Y, Z$ in the Sun-centered inertial reference frame.

It is natural to define the
symmetric and antisymmetric parts of
the $d_\mn$ coefficients,
and we do this with factors of a half:
\beq
d^\pm_{\mu\nu} \equiv \half(d_{\mu\nu}\pm d_{\nu\mu})
\, .
\eeq
To keep track of independent coefficients with
symmetric or antisymmetric pairs of indices,
we adopt the following ordering:
$TX, TY, TZ, XY, YZ, ZX$.
In addition to $d^\pm_\mn$, this affects
the first two indices of
$\gg\mu\nu\la$, and $H_\mn$.

The 24 independent components of $g_{\la\mu\nu}$
are naturally expressed in terms of
four axial components $\glA\mu$,
four trace components $\gT\mu$,
and 16 mixed-symmetry components $\gm\la\mu\nu$.
The Appendix provides the definitions of these components,
and discusses this decomposition in more detail.
Table \ref{mixed} shows
that the mixed-symmetry $g$ coefficients
are split into two independent dimension $8$
subspaces,
one with all three indices distinct,
and the other with a pair of repeated indices.
The index structure of $\gg\la\mu\nu$
is shared with the spacetime torsion tensor,
and limits on torsion have been generated
based on experiments testing Lorentz symmetry
\cite{08ktr}.

\section{Field redefinitions and observable combinations}
A variety of dependences exist between
the coefficients for Lorentz violation
in flat spacetime
\cite{97SME, 98SME}
and in gravity
\cite{akgrav}.
They can be understood through
field redefinitions,
which yield 32 coefficient combinations
in each fermion sector that are unobservable
at leading or higher orders in Lorentz violation.
This reduces
the number of coefficients appearing in equation \rf{lagr}
to a smaller set of basis combinations.

Field redefinitions have been considered
in the context of
the photon sector \cite{09Nonmin},
Finsler geometry \cite{2011finsler},
the pure-gravity sector \cite{11akjt},
supersymmetry \cite{02MBAK},
and nonlocality \cite{06RLnonlocal}.
A related topic involves
studies of particular coordinate transformations,
showing the relationship between photons and fermions
\cite{04akqb},
or between the photon coefficients and various early test models
\cite{02akmmEM}.
The absorption of the $f$ coefficient into the $c$ coefficient
has been studied in the context of the Dirac theory \cite{06Altschul}
and classical kinematics \cite{classical}.
A number of results for the fermion sector
are presented in Ref.~\cite{2002DCPM}.

\begin{table}	
\caption{\label{unobservables}Leading-order unobservable combinations of fermion-sector SME coefficients for a single particle in Minkowski space}							
\begin{indented}							
\item[]\begin{tabular}{@{}cll}
\br
{\bf Unobservable combination}  & {\bf Redefinition} & {\bf \#}\\
\mr
$	a_\mu	$&$	\ps=e^{-ia\cdot x}\ch	$&$	4	$ \\[4pt]	
$	a_\mu + m e_\mu	$&$	\ps=\left(1 + v_\mu\ga^\mu\right)\ch \, , \ v_\mu \mbox{ real}	$&$	4	$ \\[4pt]	
$	b_\mu+\glA\mu	$&$	\ps=\left(1 + v_\mu\ga_5\ga^\mu\right)\ch \, , \ v_\mu \mbox{ real}	$&$	4	$ \\[4pt]	
$	H_{\mn} + \fr 1 2 m {\ep_{\mn}}^{\si\ta} d_{\si\ta}	$&$	\ps=\left(1 + v_\mn\si^\mn\right)\ch \, , \ v_\mn \mbox{ real}	$&$	6	$ \\[4pt]	
$	c_{\mn}-c_{\nu\mu}	$&$	\ps=\left(1 - \fr 1 4 i c_\mn\si^\mn\right)\ch	$&$	6	$ \\[4pt]	
$	f_\mu	$&$	\ps=\left(1 + \fr 1 2 i f_\mu\ga_5\ga^\mu\right)\ch	$&$	4	$ \\[4pt]	
$	\gT\mu	$&$	\ps=\left(1 - \fr 1 6 i \gT\mu \ga^\mu\right)\ch	$&$	4 \ \mbox{\bf Total: } 32	$ \\	
\hlinethree
\end{tabular}							
\end{indented}							
\end{table}

Table \ref{unobservables}
summarizes the unobservable combinations of SME coefficients
for each fermion sector
in Minkowski spacetime.
The first column lists the unobservable combinations,
the second indicates the field redefinition used to show
the result,
while the last column gives the number
of independent conditions involved.
The first line
expresses the result that
the four components of the coefficient $a_\mu$
are unobservable in experiments within a single fermion sector.
This holds at all orders
in Lorentz violation \cite{97SME},
and corresponds physically to
an unobservable shift in the
energy and momentum of the system.
This differs from curved spacetime,
where the components of $a_\mu$
must vary to maintain compatibility with
the geometrical structure,
and only one component is unobservable
\cite{11akjt}.
The next six lines involve
redefinitions of the form
$\ps =(1+v\cdot \Gamma)\ch$,
where $\Gamma$ is one of
$\ga^\mu$, $\ga_5\ga^\mu$, or $\si^\mn$,
\cite{98SME,2002DCPM},
and $v$ is complex-valued with appropriate contractions.
The dependences found are valid at leading order in Lorentz violation,
and some may be valid at higher orders too.
The second, third, and fourth lines
show linear dependences between
operators of dimension 3 and 4.
In particular,
the $e_\mu$ coefficient can be absorbed into
$a_\mu$,
the four axial components of $g_{\la\mu\nu}$
can be absorbed into $b_\mu$,
and the six antisymmetic components of $d_\mn$
can be absorbed into $H_\mn$.
The unobservable combinations
listed in the
second, third and fourth lines
have
orthogonal expressions that are observable.
These can be obtained by reversing the sign between the two terms.
In the case of $H_\mn$ and $d_\mn$,
the observable combinations are:
 \bea
&&
H_{TX} + m d^-_{YZ}\, , \ \
H_{TY} + m d^-_{ZX}\, , \ \
H_{TZ} + m d^-_{XY}\, , \ \
\nonumber\\
&&
H_{YZ} - m d^-_{TX}\, , \ \
H_{ZX} - m d^-_{TY}\, , \ \
H_{XY} - m d^-_{TZ}\, .
\label{dHobservable}
\eea

In total,
Table \ref{unobservables}
lists 32 unobservable combinations
of SME coefficients.
These linear conditions reduce the number of independent
SME coefficients for each fermion sector
from the 76 appearing in  equation \rf{lagr}
to 44.
The 44 combinations may be counted as follows.
The combination $b_\mu-m \glA\mu$ has 4 components;
the symmetric and traceless $c_{\mu\nu}$ and $d_{\mu\nu}$
expressions contribute 9+9;
the combination of $H_{\mu\nu}$ with $d^-_{\mn}$
gives 6 more;
and, finally,
the mixed-symmetry components $\gm \la\mu\nu$
provide 16 independent combinations.

\section{The tilde observables in the fermion sectors}
How many of the 44 independent coefficients
that are in principle observable in each fermionic sector
after redefinitions have been accounted for,
are experimentally observable?
Based on the results of a number of analyses
of experiments with fermions,
the answer to this is that all can be accessed.

The particular combinations of SME coefficients
that appear as observables in experiments
are conventionally denoted with a tilde accent.
Forty of these can be found from
results with clock-comparison experiments
\cite{99akcl}
and space tests \cite{SpaceTests}.
Analyses with the Penning-trap system
\cite{97Penning},
which can confine both particles and antiparticles,
provide the motivation for the remaining four tilde coefficients.
Three of these are the $b$-type coefficients
relevant for antiparticles,
denoted $\bt^*_J$ with $J=X,Y,Z$.
They differ from the $\bt_J$ coefficients
in the sign of the $d$ and $H$ terms.
The remaining degree of freedom is
in the diagonal part of the $c$ coefficient.
Since $c$ is traceless,
there are three independent combinations
that can be formed on the diagonal,
two of which are the quadrupole observable $\ct_Q$,
and $\ct_-$, arising in clock-comparison experiments.
In Penning-trap systems,
the cyclotron frequency is sensitive to the combination
$c_{TT}+c_{XX}+c_{YY}$, and this is an option for defining
the 44th tilde observable.
However, due to the popularity in theoretical studies
of isotropic Lorentz violation,
the choice $\ct_{TT} := mc_{TT}$
has instead been made.
So, the basis for the diagonal
$c$ coefficients includes $\ct_Q$ and $\ct_-$
in parallel with $\dt_Q$ and $\dt_-$ for the similarly traceless $d$
coefficient;
however,
the last diagonal element for the $c$ coefficients is $\ct_{TT}$
while for $d$ it is $\dt_+$.

The full set of 44 tilde observables first appeared in Table XVII of the
{\em Data Tables for Lorentz and CPT Violation,}
January 2009 edition \cite{12Tables}.
Tables \ref{ctildetable} and \ref{tildetable}
provide the tilde definitions in a form that
differs in the order of presentation,
gives the decomposition of the $g$ coefficients
into axial, trace, and mixed-symmetry parts explicitly,
and collects together terms that are linearly dependent.

Since Lorentz violation is small,
the tilde definitions are necessarily linear,
and amount to a $44\times 44$ matrix transformation
applied to the basis of 44 independent SME coefficients
in the fermion sector.
It is of interest to investigate
the block structure of this matrix.

Table \ref{ctildetable} lists
the 9 observable $c$ combinations,
which decouple from the other observables.
The first two columns in this table,
and in Table \ref{tildetable},
express the tilde definitions
with an assignment sign understood between them.
The last column indicates the number of independent coefficients.
The first three entries of Table \ref{ctildetable}
are the observables for the diagonal.
These definitions involve
all four basic diagonal terms $c_{TT},\ c_{XX},\ c_{YY}$, and $c_{ZZ}$,
and the traceless condition can be used to eliminate any one of these.
So, the definitions of the $\ct$ coefficients
include a $3\times 3$ block mixing $c_{XX},\ c_{YY}$, and $c_{ZZ}$,
and six off-diagonal entries.
			\begin{table}							
			\caption{\label{ctildetable}Definitions of the $c$-tilde coefficients in the minimal fermion sector}							
			\begin{indented}							
			\item[]							
			\begin{tabular}{cll}							
			\hlineone							
			\multicolumn{1}{l}{{\bf Tilde combination}}	&	\multicolumn{1}{l}{{\bf Non-tilde combination}}	&	\multicolumn{1}{c}{{\bf Components}}	\\		
			\hlinetwo							
		$	\tilde{c}_Q	$&$	m(c_{XX}+c_{YY}-2c_{ZZ})	$&		\\	[2 pt]	
		$	\tilde{c}_-	$&$	m(c_{XX}-c_{YY})	$&		\\	[2 pt]	
		$	\tilde{c}_{TT}	$&$	m c_{TT}	$&	3	\\	[2 pt]	
			\hlinetwo							
		$	\tilde{c}_{TJ}	$&$	m(c_{TJ}+c_{JT})	$&	3	\\	[2 pt]	
		$	\ct_{X}	$&$	m(c_{YZ}+c_{ZY})	$&	1	\\	[2 pt]	
		$	\ct_{Y}	$&$	m(c_{XZ}+c_{ZX})	$&	1	\\	[2 pt]	
		$	\ct_{Z}	$&$	m(c_{XY}+c_{YX})	$&	1 \mbox{\bf Total: 9}	\\	[2 pt]	
			\hlinethree							
			\end{tabular}							
			\end{indented}							
			\end{table}							

Table \ref{tildetable}
gives the definitions of the remaining $35$ tilde observables.
Note that the observable combinations
in the second column,
which are orthogonal to the combinations in lines 3 and 4
of Table \ref{unobservables},
are grouped using parentheses.
The matrix implicit in this table has 8 blocks,
separated by horizontal lines:
one of dimension $6$,
three of dimension $5$,
three of dimension $4$,
and one of dimension $2$.
A bolder line separates the
first four from the last four,
and this split matches the splitting
in the mixed-symmetry $g$ components
indicated by the horizontal line in Table \ref{mixed}.
In fact,
each of the eight subspaces contains an independent
set of two of the $\gm{}{}{}$ coefficients.
In the presence of Lorentz violation in the form of $g$ coefficients only,
the first four blocks would each have dimension $4$,
because they would involve the axial $\glA{}$ as well,
and the remaining four would each have dimension $2$.
This would give access to 20 of the 24 $g$ coefficients,
with the four traceless ones being inaccessible at leading order,
consistent with the bottom line of Table \ref{unobservables}.

In addition to the two $\gm{}{}{}$ entries,
the $6\times 6$ block
contains the three diagonal entries of the $d$ coefficient,
and $b_T$,
while the three $5\times 5$ blocks
contain $b_J$, $\ep_{JKL}H_{KL}$, and the symmetric $d^+_{TJ}$ coefficients,
with one $J$ value in each.
The three $4\times 4$ blocks,
in addition to the $\gm{}{}{}$ entries,
contain $H_{TJ}$ and the remaining three $d^+$ coefficients,
again with one $J$ value in each.

			\begin{table}							
			\caption{\label{tildetable}Definitions of the minimal-fermion-sector tilde coefficients, excluding the $c$'s.}							
			\begin{indented}							
			\item[]							
			\begin{tabular}{cll}							
			\hlineone							
			\multicolumn{1}{l}{{\bf Tilde combination}}	&	\multicolumn{1}{l}{{\bf Non-tilde combination}}	&	\multicolumn{1}{l}{{\bf Components}}	\\		
			\hlinetwo							
		$	\bt_T	$&$	(b_T - m \glA T) + m \gm X Y Z	$&		\\	[2pt]	
		$	\gt_T	$&$	(b_T - m \glA T) - 2 m \gm X Y Z	$&		\\	[2pt]	
		$	\dt_Q	$&$	m(d_{XX}+d_{YY}-2 d_{ZZ}) + 3 m \gm X Y Z	$&		\\	[2pt]	
		$	\dt_+	$&$	m(d_{XX}+d_{YY})	$&		\\	[2pt]	
		$	\dt_-	$&$	m(d_{XX}-d_{YY})	$&		\\	[2pt]	
		$	\gt_c	$&$	2 m \gm X Y Z + m \gm Y Z X	$&	6	\\	[2pt]	
			\hlinetwo							
		$	\bt_X^*	$&$	(b_X - m \glA X) + (H_{YZ} - m d^-_{TX}) + m d^+_{TX} + m \gm Y Z T	$&		\\	[2pt]	
		$	\bt_X	$&$	(b_X - m \glA X) - (H_{YZ} - m d^-_{TX}) - m d^+_{TX} + m \gm Y Z T	$&		\\	[2pt]	
		$	\dt_X	$&$	 - \frac 1 2 (H_{YZ}-m d^-_{TX}) + \frac 3 2 m d^+_{TX}	$&		\\	[2pt]	
		$	\gt_{DX}	$&$	-(b_X - m \glA X) + 2 m \gm Y Z T	$&		\\	[2pt]	
		$	\gt_{TX}	$&$	-m \gm Y Z T - 2 m \gm T Y Z	$&	5	\\	[2pt]	
			\hlinetwo							
		$	\bt_Y^*	$&$	(b_Y - m \glA Y) + (H_{ZX} - m d^-_{TY}) + m d^+_{TY} + m \gm Z X T	$&		\\	[2pt]	
		$	\bt_Y	$&$	(b_Y - m \glA Y) - (H_{ZX} - m d^-_{TY}) - m d^+_{TY} + m \gm Z X T	$&		\\	[2pt]	
		$	\dt_Y	$&$	 - \frac 1 2 (H_{ZX} - m d^-_{TY}) + \frac 3 2 m d^+_{TY}	$&		\\	[2pt]	
		$	\gt_{DY}	$&$	-(b_Y - m \glA Y) + 2 m \gm Z X T	$&		\\	[2pt]	
		$	\gt_{TY}	$&$	-m \gm Z X T - 2 m \gm T Z X	$&	5	\\	[2pt]	
			\hlinetwo							
		$	\bt_Z^*	$&$	(b_Z - m \glA Z) + (H_{XY} - m d^-_{TZ}) + m d^+_{TZ} + m \gm X Y T	$&		\\	[2pt]	
		$	\bt_Z	$&$	(b_Z - m \glA Z) - (H_{XY} - m d^-_{TZ}) - m d^+_{TZ} + m \gm X Y T	$&		\\	[2pt]	
		$	\dt_Z	$&$	 - \frac 1 2 (H_{XY} - m d^-_{TZ}) + \frac 3 2 m d^+_{TZ}	$&		\\	[2pt]	
		$	\gt_{DZ}	$&$	-(b_Z - m \glA Z) + 2 m \gm X Y T	$&		\\	[2pt]	
		$	\gt_{TZ}	$&$	-m \gm X Y T - 2 m \gm T X Y	$&	5	\\	[2pt]	
			\hlinethree							
		$	\Ht_{TX}	$&$	(H_{TX} + m d^-_{YZ}) - m d^+_{YZ} - m \gm T X T + m \gm X Y Y  	$&		\\	[2pt]	
		$	\dt_{YZ}	$&$	2 m d^+_{YZ} - m \gm T X T - 2 m \gm X Y Y	$&		\\	[2pt]	
		$	\gt_{XZ}	$&$	- 2 m \gm T X T - m \gm X Y Y	$&		\\	[2pt]	
		$	\gt_{XY}	$&$	-m \gm T X T + m \gm X Y Y	$&	4	\\	[2pt]	
			\hlinetwo							
		$	\Ht_{TY}	$&$	(H_{TY} + m d^-_{ZX}) - m d^+_{ZX}  - m \gm T Y T + m \gm Y Z Z	$&		\\	[2pt]	
		$	\dt_{ZX}	$&$	2 m d^+_{ZX} - m \gm T Y T - 2 m \gm Y Z Z	$&		\\	[2pt]	
		$	\gt_{YX}	$&$	- 2 m \gm T Y T - m \gm Y Z Z	$&		\\	[2pt]	
		$	\gt_{YZ}	$&$	-m \gm T Y T + m \gm Y Z Z	$&	4	\\	[2pt]	
			\hlinetwo							
		$	\Ht_{TZ}	$&$	(H_{TZ} + m d^-_{XY}) - m d^+_{XY} - m \gm T Z T + m \gm Z X X	$&		\\	[2pt]	
		$	\dt_{XY}	$&$	2 m d^+_{XY} - m \gm T Z T - 2 m \gm Z X X	$&		\\	[2pt]	
		$	\gt_{ZY}	$&$	- 2 m \gm T Z T - m \gm Z X X	$&		\\	[2pt]	
		$	\gt_{ZX}	$&$	-m \gm T Z T + m \gm Z X X	$&	4	\\	[2pt]	
			\hlinetwo							
		$	\gt_-	$&$	-m \gm T X X + m \gm T Y Y	$&		\\	[2pt]	
		$	\gt_Q	$&$	-3m \gm T X X - 3 m \gm T Y Y	$&	2 \ \ {\bf Total: 35}	\\	[2pt]	
			\hlinethree							
			\end{tabular}							
			\end{indented}							
			\end{table}							

Reference \cite{12Tables}
provides experimental measurements of the tilde observables
for the electron, proton, and neutron sectors,
as well as a summary table giving the maximal experimental sensitivity
to each coefficient with the assumption that other coefficients don't contribute.
Sensitivity to all nine $\ct$ observables
exists for the electron and proton sectors,
and six of them in the neutron sector.
No experimental sensitivities have been published
to tilde coefficients in the $2\times 2$ block
in any of the three fermion sectors.
Among the $5\times 5$ blocks,
no sensitivities to the $J=Z$ set have been published for protons or neutrons.
The least explored fermion sector is that of the proton,
for which only six sensitivities
exist outside of the $c$ block:
these are within the $J=X$ and $J=Y$
$5\times 5$ blocks.
Experimental access to the $6\times 6$ block
can be attained by taking into account the boost of the
experiment relative to the standard reference frame.
This has been done for the electron sector
using a torsion-pendulum experiment
\cite{08HeckelPRD},
and for the neutron sector using a dual maser system
\cite{04Cane}.
At present, there are no published sensitivities to coefficients
in the $6\times 6$ block of the proton sector.

\section{Inverting the tilde definitions}
In some experiments,
measurements have been made of SME coefficients
that differ from the tilde observables.
This has been done by decoupling the observables
through combining independent results.
For results such as these,
the sensitivities to the tilde observables
can be deduced by using the inverse of the tilde definitions.

The linear transformation relating
the 44 independent SME coefficient combinations
and the 44 tilde observables
is invertible,
and we present the relevant expressions in
Tables \ref{c:table} and
\ref{Inversetildetable}.
Table \ref{c:table} expresses
the inverse of the $9\times 9$ matrix
defining the $\ct$  observables
in Table \ref{ctildetable}.
In this inverse,
the traceless condition
$
c_{TT} = c_{XX} + c_{YY} + c_{ZZ}
$
has been imposed to eliminate $c_{TT}$,
and this may be verified by
adding the first three lines of the table.
	\begin{table}								
	\caption{\label{c:table} Minimal-fermion-sector $c$ coefficients in terms of tilde quantities.}								
	\begin{indented}								
	\item[]								
		\begin{tabular}{cll}							
		\hlineone							
		{\bf Non-tilde combination}	& \multicolumn{1}{l}{{\bf Tilde combination}}	& \multicolumn{1}{l}{{\bf Components}}\\					
		\hlinetwo							
	$	m c_{XX} 	$&$	 \frac 1 6 (\ct_Q + 3 \ct_-  + 2 \ct_{TT})	$&		\\	[2 pt]	
	$	m c_{YY} 	$&$	 \frac 1 6 (\ct_Q -3 \ct_-  + 2 \ct_{TT})	$&		\\	[2 pt]	
	$	m c_{ZZ} 	$&$	 \frac 1 3 (-\ct_Q + \ct_{TT})	$&	3	\\	[2 pt]	
		\hlinetwo							
	$	m \abs{\ve_{JKL}} c_{KL} 	$&$	 \ct_J	$&	1+1+1	\\	[2 pt]	
	$	m(c_{TJ}+c_{JT}) 	$&$	 \ct_{TJ}	$&	1+1+1 \ \ {\bf Total: 9}	\\	[2 pt]	
		\hlinethree							
	\end{tabular}								
	\end{indented}								
	\end{table}								

Table \ref{Inversetildetable}
expresses the inverse of the remaining $35$ tilde definitions .
The block structure has been carried over from Table \ref{tildetable}
and horizontal lines are again used to separate the 8 subspaces.
In the $6\times 6$ block
$d_{TT}$ has been eliminated using the condition
that $d_\mn$ is traceless,
$d_{TT} = d_{XX} + d_{YY} + d_{ZZ}$.
The next three sections of Table~\ref{Inversetildetable}
are $5 \times 5$ blocks based on the definitions
for
$\bt_J, \, \bt_J^*, \, \dt_J, \, \gt_{DJ}, \, \gt_{TJ}$,
with $J=X, Y, Z$.

In each of the first four sections of Table~\ref{Inversetildetable},
the two mixed-symmetry components
satisfy one of the four nonaxial conditions \rf{nonaxial}.
So, for example, an expression for $m \gm Z X Y$ can be found by adding the
last two lines in the first section of the Table,
giving
\beq
m \gm Z X Y = \frac 1 3 (\bt_T - \gt_T) - \gt_c
\, .
\eeq

In each of the bottom four blocks in Table \ref{Inversetildetable},
the two mixed-symmetry coefficients $\gm{}{}{}$
satisfy a traceless condition \rf{traceless}.
The four traceless conditions are given explicitly
in Table \ref{mixed}.
Adding the last two lines of Table \ref{Inversetildetable},
and noting the last line of Table \ref{mixed},
one can verify that
\beq
m \gm T Z Z = \frac 1 3 \gt_Q
\, .
\eeq

		\begin{table}							
		\caption{\label{Inversetildetable}Minimal-fermion-sector SME coefficients in terms of tilde quantities (excluding $c$'s).}							
		\begin{indented}							
		\item[]							
		\begin{tabular}{lll}							
		\hlineone							
		{\bf Non-tilde combination}	&	\multicolumn{1}{l}{{\bf Tilde combination}}	&	\multicolumn{1}{l}{{\bf Components}}	\\		
		\hlinetwo							
	$	b_T - m \glA T 	$&$	 \frac 1 3 (2 \bt_T + \gt_T)	$&		\\	[2pt]	
	$	m d_{XX} 	$&$	 \frac 1 2 (\dt_+ + \dt_- )	$&		\\	[2pt]	
	$	m d_{YY} 	$&$	 \frac 1 2 (\dt_+ -\dt_- )	$&		\\	[2pt]	
	$	m d_{ZZ} 	$&$	 \frac 1 2 (\bt_T  - \gt_T + \dt_+ -\dt_Q)	$&		\\	[2pt]	
	$	m \gm X Y Z 	$&$	 \frac 1 3 (\bt_T - \gt_T)	$&		\\	[2pt]	
	$	m \gm Y Z X 	$&$	 - \frac 2 3 (\bt_T - \gt_T) +  \gt_c	$&	6	\\	[2pt]	
		\hlinetwo							
	$	b_X - m\glA X 	$&$	  \frac 1 3 (\bt_X + \bt^*_X -\gt_{DX})	$&		\\	[2pt]	
	$	H_{YZ} - m d^-_{TX} 	$&$	  - \frac 3 8 (\bt_X - \bt^*_X) - \frac 1 2 \dt_X	$&		\\	[2pt]	
	$	m d^+_{TX}	$&$	  - \frac 1 8 (\bt_X - \bt^*_X) + \frac 1 2 \dt_X	$&		\\	[2pt]	
	$	m\gm Y Z T 	$&$	 \frac 1 6 (\bt_X + \bt^*_X +  2 \gt_{DX})	$&		\\	[2pt]	
	$	m\gm T Y Z 	$&$	 - \frac 1 {12} (\bt_X + \bt^*_X +  2 \gt_{DX} + 6 \gt_{TX})	$&	5	\\	[2pt]	
		\hlinetwo							
	$	b_Y - m \glA Y	$&$	 \frac 1 3 (\bt_Y + \bt^*_Y -\gt_{DY})	$&		\\	[2pt]	
	$	H_{ZX} - m d^-_{TY} 	$&$	 -\frac 3 8 (\bt_Y -  \bt^*_Y) - \frac 1 2  \dt_Y	$&		\\	[2pt]	
	$	m d^+_{TY}	$&$	  -\frac 1 8 (\bt_Y - \bt^*_Y) + \frac 1 2 \dt_Y	$&		\\	[2pt]	
	$	m \gm Z X T 	$&$	 \frac 1 6 (\bt_Y + \bt^*_Y+ 2 \gt_{DY})	$&		\\	[2pt]	
	$	m \gm T Z X 	$&$	 - \frac 1 {12}(\bt_Y + \bt^*_Y + 2 \gt_{DY} + 6 \gt_{TY})	$&	5	\\	[2pt]	
		\hlinetwo							
	$	b_Z - m \glA Z 	$&$	 \frac 1 3 (\bt_Z + \bt^*_Z - \gt_{DZ})	$&		\\	[2pt]	
	$	H_{XY} - m d^-_{TZ}	$&$	 - \frac 3 8 (\bt_Z - \bt^*_Z) - \frac 1 2 \dt_Z	$&		\\	[2pt]	
	$	m d^+_{TZ}	$&$	 - \frac 1 8 (\bt_Z - \bt^*_Z) + \frac 1 2 \dt_Z	$&		\\	[2pt]	
	$	m \gm X Y T 	$&$	 \frac 1 6 (\bt_Z + \bt^*_Z + 2 \gt_{DZ})	$&		\\	[2pt]	
	$	m \gm T X Y 	$&$	 - \frac 1 {12} (\bt_Z + \bt^*_Z + 2 \gt_{DZ} +6 \gt_{TZ})	$&	5	\\	[2pt]	
		\hlinethree							
	$	H_{TX} + m d^-_{YZ} 	$&$	 \Ht_{TX} + \half (\dt_{YZ} - \gt_{XY} - \gt_{XZ})	$&		\\	[2pt]	
	$	m d^+_{YZ}	$&$	\frac 1 2 (\dt_{YZ} + \gt_{XY} - \gt_{XZ})	$&		\\	[2pt]	
	$	m \gm T X T 	$&$	 - \frac 1 3 (\gt_{XY} + \gt_{XZ})	$&		\\	[2pt]	
	$	m \gm X Y Y 	$&$	 \frac 1 3 (2\gt_{XY} - \gt_{XZ} )	$&	4	\\	[2pt]	
		\hlinetwo							
	$	H_{TY} + m d^-_{ZX}	$&$	 \Ht_{TY} + \half (\dt_{ZX} - \gt_{YZ} - \gt_{YX})	$&		\\	[2pt]	
	$	m d^+_{ZX}	$&$	\frac 1 2 (\dt_{ZX} + \gt_{YZ} - \gt_{YX})	$&		\\	[2pt]	
	$	m \gm T Y T 	$&$	 - \frac 1 3 (\gt_{YZ} + \gt_{YX} )	$&		\\	[2pt]	
	$	m \gm Y Z Z	$&$	 \frac 1 3 (2\gt_{YZ} - \gt_{YX} )	$&	4	\\	[2pt]	
		\hlinetwo							
	$	H_{TZ} + m d^-_{XY}	$&$	 \Ht_{TZ} + \half (\dt_{XY} - \gt_{ZX} - \gt_{ZY})	$&		\\	[2pt]	
	$	m d^+_{XY}	$&$	\frac 1 2 (\dt_{XY} + \gt_{ZX} - \gt_{ZY})	$&		\\	[2pt]	
	$	m \gm T Z T 	$&$	 - \frac 1 3 (\gt_{ZX} + \gt_{ZY})	$&		\\	[2pt]	
	$	m \gm Z X X 	$&$	 \frac 1 3 (2\gt_{ZX} -\gt_{ZY} )	$&	4	\\	[2pt]	
		\hlinetwo							
	$	m \gm T X X 	$&$	 - \frac 1 6 (3 \gt_- + \gt_Q)	$&		\\	[2pt]	
	$	m \gm T Y Y 	$&$	  \frac 1 6 (3 \gt_- - \gt_Q)	$&	2 \ \ {\bf Total: 35}	\\	[2pt]	
		\hlinethree							
		\end{tabular}							
		\end{indented}							
		\end{table}							

\section{Limits on SME observables}
We next use expressions in Table \ref{Inversetildetable} to estimate
limits on several of the electron-sector tilde observables.

The three expressions
\beq
b_J - m\glA J = \frac 1 3 (\bt_J + \bt^*_J -\gt_{DJ})
\, ,
\label{bexpr}
\eeq
appearing as the first line in each
of the dimension 5 blocks,
generate limits on $b^*_J$ and $\gt_{DJ}$
from the existing limits
on $b_J$, for $J=1,2,3$.
In this case,
the bounds are
$|b_J| < 50$~rad/sec
$\simeq 3.3 \times 10^{-23}$ GeV,
based on Penning-trap experiments
\cite{99Dehmelt}.
Setting the axial components $\glA J$ to zero,
since the $g$ coefficients were not included
in analyses when these results were published,
we may deduce separate order-of-magnitude limits on each of the tilde
coefficients on the right-hand side of Eq.\ \rf{bexpr}.
To do this,
we assume that the limit on $|b_J|$ is at the $2\si$ level;
then,
considering each tilde coefficient
in isolation,
we deduce that
\beq
|\bt^{*e}_J|, \, |\gt_{DJ}^e| < 9.9\times 10^{-23} \gev\, \simeq 10^{-22} \gev \, .
\label{resultbg}
\eeq
The same bound also follows for $|\bt_J|$,
but is not competitive with limits from other experiments.

Several limits on electron-sector tilde combinations follow from
inverse Compton scattering bounds on SME coefficients
\cite{07AltschulAstrophys}.
Following a similar argument as above,
the limit
$|d_{YZ}+d_{ZY}|< 7\times 10^{-15}$
may be inserted into the second line of the first dimension-four block
in Table \ref{Inversetildetable}.
Using the electron mass $m=0.51\times 10^{-3} \gev$
yields
\beq
|\gt_{XY}^e| \, , |\gt_{XZ}^e| < 10^{-17} \gev
\, .
\label{resultgg1}
\eeq
The second lines of the other two dimension-four blocks
in Table \ref{Inversetildetable},
combined with the results
$|d_{XZ}+d_{ZX}|< 2\times 10^{-14}$
and
$|d_{XY}+d_{YX}|< 2\times 10^{-15}$
in \cite{07AltschulAstrophys},
lead to
\bea
|\gt_{YZ}^e| \, , |\gt_{YX}^e| &<& 10^{-17} \gev
\, , \label{resultgg2}
\\
|\gt_{ZX}^e| \, , |\gt_{ZY}^e| &<& 10^{-18} \gev
\, , \label{resultgg3}
\eea
respectively.

A limit on the electron-sector $\dt_Z$ can be extracted from
the result $|d_{TZ}| < 8\times 10^{-17}$
in Ref.~\cite{07AltschulAstrophys}.
In that analysis,
$H_{XY}$ may be interpreted as having been
absorbed into the antisymmetric part of
$m d_{TZ}$.
By reinstating it via the observable
combination seen in equation \rf{dHobservable},
the quantity bounded becomes
$
m d_{TZ}
\rightarrow m d^+_{TZ} + m d^-_{TZ} - H_{XY}
$.
This can be expanded in terms of tilde expressions
using the results in
the $Z$ component $5\times 5$ block of Table~\ref{Inversetildetable},
giving
$
m d_{TZ}
\rightarrow (\bt_Z - \bt^*_Z)/4 + \dt_Z
$.
While the resulting sensitivities to $\bt_Z$ and $\bt_Z^*$
are weaker than ones derived from other systems,
the sensitivity to $\dt_Z$
is the only one known at present.
The latter result is
\beq
|\dt_Z^e| < 10^{-19} \gev
\, .
\label{resultdZ}
\eeq
We estimate a 90\% confidence level
for each of these astrophysical limits.
The electron-sector bounds extracted here from
Penning-trap experiments and astrophysical data
are consistent
with the estimated maximal sensitivities
appearing in Ref. \cite{12Tables},
but have not been described in the literature to date.

\section{Summary and discussion}
We have considered the algebraic structure
of the coefficients for Lorentz violation
in the flat-space fermion sector of the minimal SME.
This has revealed a block structure in the system of experimental
observables.
By inverting the definitions,
thirteen limits on the tilde observables
in the electron-positron sector have been found,
and are given in equations
\rf{resultbg},
\rf{resultgg1},
\rf{resultgg2},
\rf{resultgg3},
and
\rf{resultdZ}.

To date,
most Lorentz tests
have concentrated on seeking sidereal variations in signals,
due to the rotation of the laboratory relative to
the standard Sun-centered inertial reference frame.
When the linear motion of the system
relative to the inertial reference frame
is considered
modifications involving the boost factor
$\ga = (1-v^2)^{-1/2}$
in the expressions for the $44$
fermion observables can be expected.
This has been done in
Lorentz tests
with a spin-polarized torsion pendulum
\cite{08HeckelPRD}
and with a dual maser system
\cite{04Cane},
as mentioned earlier.
In Lorentz tests where the particles are
significantly boosted
relative to the inertial reference frame,
such as in muon experiments
\cite{08Muon},
the observables will necessarily involve
factors of $\ga$.
In such cases,
the relativistic hamiltonian
\cite{99akclNonRel}
is needed,
and the $\bt_J$ and $\bt^*_J$
observables take the modified
`\hacek'
form
\cite{00BKL:muonPRL}:
\beq
\check b_X^\mp := \fr 1 \ga (b_X - m\glA X + m\gm Y Z T)
    \mp (H_{YZ}-m d^-_{TX}) \mp m d^+_{TX}
    \, ,
\eeq
with $\check b^\mp_Y$ and $\check b^\mp_Z$ defined by cyclic rotations of
$X,Y,Z$.
The \hacek\ expressions equal the tilde coefficients
in the $\ga \rightarrow 1$ nonrelativistic limit.

Another way to gain sensitivities
to SME coefficients is to
take interactions into account.
This has been done in the context of
relativistic nuclear binding effects
in atomic clocks
\cite{09AltschulClocks}.
The results indicate that,
under appropriate circumstances,
separation can be achieved between
the $b$ and $g$ coefficients,
and between the $H$ and $d$ coefficients.

As noted earlier,
the $6\times 6$ block of tilde observables
for the proton sector of Table \ref{tildetable}
has no published bounds on it.
The boosted observable expressions mix the spatial components
from the three $5\times 5$ blocks
with the time components in the $6\times 6$ block.
It follows that any experiments sensitive to
tilde coefficients in the proton sector
have the potential to access
the $6\times 6$ block in Table \ref{tildetable}.
In addition,
the absence of any sensitivities to
one of the $5\times 5$ blocks
in the proton sector,
and all three of the $4\times 4$ blocks,
means Lorentz tests
have the potential to yield a rich crop of new results.

\section*{Acknowledgement}
The hospitality of the Indiana University Center
for Spacetime Symmetries is gratefully acknowledged.

\appendix
\section{Irreducible components of $\gg \la \mu \nu$}
The SME coefficients $g_{\mu\nu\al}$ are antisymmetric in the first two indices,
which means they comprise 24 independent numbers.
This observer tensor can be uniquely decomposed into three standard irreducible components,
the axial part $\glA {\mn\al}$,
the trace part $\gT {\mn\al}$,
and the mixed-symmetry part $\gm \mu\nu\al$:
\beq
g_{\mn\la} = \glA{\mn\la} + \gT{\mn\la} + \gm \mu\nu\la \, .
\eeq
Expressions defining each of these components are as follows:
\bea
\glA{\mu\nu\la} &=& \frac 1 6 \gg\si\ka\ta \ep^{\si\ka\ta\al} \ep_{\al\mu\nu\la}
   = \frac 1 3 (g_{\mu\nu\la} + g_{\la\mu\nu}+ g_{\nu\la\mu})
\label{def:axial} \, , \\
\gT{\mu\nu\la} &=& \frac 1 3 \et^{\al\be}(g_{\mu\al\be}\et_{\nu\la} - g_{\nu\al\be}\et_{\mu\la})
\label{def:trace}\, , \\
\gm \mu\nu\la &=& \frac 1 3 (g_{\mu\nu\la} + g_{\mu\la\nu} + \et_{\mu\la} g_{\nu\al\be} \et^{\al\be})
  -  (\mu \leftrightarrow \nu)
\label{def:mixed} \, .
\eea
The numerical values of these lower-index expressions
are independent of the choice of spacetime-metric signature
and,
separately,
of the sign convention for $\ep^{0123}$,
since the metric and the antisymmetric tensor appear quadratically.

There are four independent axial components of $\glA{\mn\la}$,
and they can be found by contraction with the antisymmetric tensor.
Hence we define a single-index axial vector
\beq
\gA \al \equiv \frac 1 6 \gg\si\ka\ta \ep^{\si\ka\ta\al}
\label{axoneindex}\, .
\eeq
Similar contractions over the three indices of
$\gT{\mn\la}$ or $\gm\mu\nu\la $ with the antisymmetric tensor
lead to a zero result.

The top half of
Table \ref{axialandtrace} lists
the axial components of $\gg\mu\nu\la$
in a reference frame with time and space coordinates $T,X,Y,Z$.
The expressions involving all but the fourth column
are valid under all sign conventions
for the spacetime metric and the totally antisymmetric tensor,
because,
in equation \rf{def:axial},
the former does not appear
and the latter appears quadratically.
The expressions involving
the lower-index $\glA\mu$
are valid for
sign conventions
$\et_{00}=+1$ and $\ep^{0123}=+1$,
or, alternatively,
$\et_{00}=-1$ and $\ep^{0123}=-1$.
This follows because,
in equation \rf{axoneindex},
one factor of the totally antisymmetric tensor appears,
and one factor of the metric is needed to
lower the index.
Thus,
the product $\et_{00} \ep^{0123}$ has to have a fixed sign.

There are four independent trace components $\gT{\mu\nu\la}$,
and they can be found by contracting the last two indices of $g_{\mu\nu\la}$
with the metric.
So, a single-index trace component is defined by
\beq
\gT \mu \equiv \gg\mu\al\be \et^{\al\be}
= {g_{\mu\al}}^\al
\label{troneindex} \, .
\eeq
This contraction gives a zero result
if applied to
$\glA{\mu\nu\la}$ or $\gm\mu\nu\la$.

The bottom half of Table \ref{axialandtrace}
lists the trace components of
$\gg\mu\nu\al$ in the standard Sun-centered frame.
All the expressions involving three-index quantities
hold under any sign conventions for $\et_{00}$
and $\ep^{0123}$,
because
equation \rf{def:trace}
involves the metric quadratically
and does not involve the totally antisymmetric tensor.
Expressions for the single-index quantities
$\guT\mu$
also hold under any sign convention
because the metric appears quadratically:
once in equation \rf{troneindex},
and once more to raise the index position.

\begin{table}
\caption{\label{axialandtrace}Axial and trace components of $\gg\mu\nu\la$.
The signs for the single-index components $\glA\nu$ hold for conventions with product $\et_{00}\, \ep^{0123}=+1$.
All other signs hold in any convention.}
\begin{indented}
\item[]\begin{tabular}{@{}l@{\hspace{22pt}}l@{\hspace{22pt}}l@{\hspace{22pt}}ll}
\br
\multicolumn{4}{@{}l}{\bf Equal triple- and single-index components} & {{\bf Expression}}\\
\mr
$	\ggA X Y Z 	$&$	 \ggA Y Z X 	$&$	 \ggA Z X Y 	$&$	 - \glA T  	$&$	 \frac 1 3 (\gg X Y Z + \gg Y Z X + \gg Z X Y)	$ \\[2pt]
$	\ggA T Y Z 	$&$	 \ggA Y Z T 	$&$	 - \ggA T Z Y 	$&$	 - \glA X 	$&$	 \frac 1 3 (\gg T Y Z + \gg Y Z T - \gg T Z Y)	$ \\[2pt]
$	\ggA T Z X 	$&$	 \ggA Z X T 	$&$	 - \ggA T X Z 	$&$	 - \glA Y 	$&$	 \frac 1 3 (\gg T Z X + \gg Z X T - \gg T X Z)	$ \\[2pt]
$	\ggA T X Y 	$&$	 \ggA X Y T 	$&$	 - \ggA T Y X 	$&$	 - \glA Z 	$&$	 \frac 1 3 (\gg T X Y + \gg X Y T - \gg T Y X)	$ \\[2pt]
\mr										
$	\ggT T X X 	$&$	 \ggT T Y Y 	$&$	 \ggT T Z Z 	$&$	 - \frac 1 3 \guT T 	$&$	 \frac 1 3 (\gg T X X + \gg T Y Y + \gg T Z Z)	$ \\[2pt]
$	\ggT T X T 	$&$	 \ggT X Y Y 	$&$	 - \ggT Z X Z 	$&$	 \frac 1 3 \guT X 	$&$	 \frac 1 3 (\gg T X T + \gg X Y Y - \gg Z X Z)	$ \\[2pt]
$	\ggT T Y T 	$&$	 \ggT Y Z Z 	$&$	 - \ggT X Y X 	$&$	 \frac 1 3 \guT Y 	$&$	 \frac 1 3 (\gg T Y T + \gg Y Z Z - \gg X Y X)	$ \\[2pt]
$	\ggT T Z T 	$&$	 \ggT Z X X 	$&$	 - \ggT Y Z Y 	$&$	 \frac 1 3 \guT Z 	$&$	 \frac 1 3 (\gg T Z T + \gg Z X X - \gg Y Z Y)	$ \\[2pt]
\br
\end{tabular}
\end{indented}
\end{table}

Table \ref{mixed}
lists the mixed-symmetry components of $\gg\mu\nu\la$
in the standard Sun-centered frame.
Since the spacetime metric appears
quadratically in equation \rf{def:mixed},
and the totally antisymmetric tensor
does not appear at all,
the expressions in Table \ref{mixed}
are valid regardless of the sign convention.

By definition,
the mixed-symmetry components $\gm\mu\nu\la$
are constrained by four conditions
that ensure they are nonaxial:
\beq
\gm\mu\nu\la \ep^{\mu\nu\la\al} = 0
\label{nonaxial}
\, ,
\eeq
and by another four conditions
that ensure they are traceless:
\beq
\gm\mu\al\be \et^{\al\be} = 0
\label{traceless}
\, .
\eeq
The four nonaxial conditions
appear explicitly
in the upper part of the first column
of Table \ref{mixed}.
The four traceless conditions
are given explicitly in the lower portion
of the first column.
These 8 linear relations among the 24 components of $\gm JKL$
ensure that the dimension of the mixed-symmetry vector space is 16.

We note that the upper half of the table
consists of components with three distinct indices,
while the lower part has components with a pair of identical indices.
It follows that the mixed-symmetry tensors
are split into two independent vector spaces,
each of dimension 8.

\begin{table}
\caption{\label{mixed}Mixed-symmetry components of $\gg\mu\nu\la$.

The signs hold regardless of the convention for $\et_{00}$ and  $\ep^{0123}$.}
\begin{indented}
\item[]\begin{tabular}{@{}ll}
\br
{\bf Components} & {{\bf Expression}}\\
\mr
$	\gm X Y Z 			$&$	 \frac 1 3 (2 \gg X Y Z - \gg Y Z X - \gg  Z X Y)	$ \\[2pt]
$	\gm Y Z X 			$&$	 \frac 1 3 (2\gg Y Z X - \gg Z X Y - \gg X Y Z)	$ \\[2pt]
$	\gm Z X Y 	=	 - \gm X Y Z - \gm Y Z X 	$&$	 \frac 1 3 (2 \gg Z X Y - \gg X Y Z - \gg Y Z X)	$ \\[2pt]
$	\gm Y Z T 			$&$	 \frac 1 3 (2\gg Y Z T + \gg T Z Y - \gg T Y Z)	$ \\[2pt]
$	\gm T Y Z 			$&$	 \frac 1 3 (2\gg T Y Z + \gg T Z Y - \gg Y Z T)	$ \\[2pt]
$	\gm T Z Y 	=	 \gm Y Z T + \gm T Y Z 	$&$	 \frac 1 3 (2 \gg T Z Y + \gg T Y Z + \gg Y Z T)	$ \\[2pt]
$	\gm Z X T 			$&$	 \frac 1 3 (2 \gg Z X T + \gg T X Z - \gg T Z X)	$ \\[2pt]
$	\gm T Z X 			$&$	 \frac 1 3 (2 \gg T Z X + \gg T X Z - \gg Z X T)	$ \\[2pt]
$	\gm T X Z 	=	 \gm Z X T + \gm T Z X 	$&$	 \frac 1 3 (2\gg T X Z + \gg T Z X + \gg Z X T)	$ \\[2pt]
$	\gm X Y T 			$&$	 \frac 1 3 (2 \gg X Y T + \gg T Y X - \gg T X Y)	$ \\[2pt]
$	\gm T X Y 			$&$	 \frac 1 3 (2 \gg T X Y + \gg T Y X - \gg X Y T)	$ \\[2pt]
$	\gm T Y X 	=	 \gm X Y T + \gm T X Y 	$&$	 \frac 1 3 (2 \gg T Y X + \gg T X Y + \gg X Y T)	$ \\[2pt]
\mr
$	\gm T X T 			$&$	 \frac 1 3 (2\gg T X T - \gg X Y Y + \gg Z X Z)	$ \\[2pt]
$	\gm X Y Y 			$&$	 \frac 1 3 (2 \gg X Y Y - \gg T X T + \gg Z X Z)	$ \\[2pt]
$	\gm Z X Z = \gm T X T + \gm X Y Y 			$&$	 \frac 1 3 (2\gg Z X Z + \gg T X T + \gg X Y Y)	$ \\[2pt]
$	\gm T Y T 			$&$	 \frac 1 3 (2 \gg T Y T + \gg X Y X - \gg Y Z Z)	$ \\[2pt]
$	\gm Y Z Z 			$&$	 \frac 1 3 (2 \gg Y Z Z - \gg T Y T + \gg X Y X)	$ \\[2pt]
$	\gm X Y X = \gm T Y T + \gm Y Z Z 			$&$	 \frac 1 3 (2 \gg X Y X + \gg T Y T + \gg Y Z Z)	$ \\[2pt]
$	\gm T Z T 			$&$	 \frac 1 3 (2 \gg T Z T - \gg Z X X + \gg Y Z Y)	$ \\[2pt]
$	\gm Z X X 			$&$	 \frac 1 3 (2 \gg Z X X - \gg T Z T + \gg Y Z Y)	$ \\[2pt]
$	\gm Y Z Y = \gm T Z T + \gm Z X X 			$&$	 \frac 1 3 (2 \gg Y Z Y + \gg T Z T + \gg Z X X)	$ \\[2pt]
$	\gm T X X 			$&$	 \frac 1 3 (2 \gg T X X - \gg T Y Y - \gg T Z Z)	$ \\[2pt]
$	\gm T Y Y 			$&$	 \frac 1 3 (2 \gg T Y Y - \gg T X X - \gg T Z Z)	$ \\[2pt]
$	\gm T Z Z = - \gm T X X - \gm T Y Y 			$&$	 \frac 1 3 (2 \gg T Z Z - \gg T X X - \gg T Y Y)	$ \\[2pt]
\br
\end{tabular}
\end{indented}
\end{table}

\end{document}